\documentclass[%
 reprint,
 amsmath,amssymb,
 aps,
]{revtex4-2}

\usepackage{xcolor}

\usepackage{graphicx}% Include figure files
\usepackage{dcolumn}% Align table columns on decimal point
\usepackage{bm}% bold math
\begin{document}

\title{Energy diffusion and prethermalization in chaotic billiards under rapid periodic driving}

\author{Wade Hodson}
\email{whodson@umd.edu}

\affiliation{Department of Physics, University of Maryland, College Park, MD 20742}

\author{Christopher Jarzynski}
\email{cjarzyns@umd.edu}

\affiliation{Institute for Physical Science and Technology,
Department of Chemistry and Biochemistry, and Department of Physics, University of Maryland, College Park, MD 20742}

\date{\today}

%TC:ignore

\begin{abstract}
    We study the energy dynamics of a particle in a billiard subject to a rapid periodic drive. In the regime of large driving frequencies $\omega$, we find that the particle's energy evolves diffusively, which suggests that the particle's energy distribution $\eta (E,t)$ satisfies a Fokker-Planck equation. We calculate the rates of energy absorption and diffusion associated with this equation, finding that these rates are proportional to $\omega^{-2}$ for large $\omega$.    Our analysis suggests three phases of energy evolution: Prethermalization on short timescales, then slow energy absorption in accordance with the Fokker-Planck equation, and finally a breakdown of the rapid driving assumption for large energies and high particle speeds.  We also present numerical simulations of the evolution of a rapidly driven billiard particle, which corroborate our theoretical results.

\end{abstract}

%TC:endignore

\maketitle

\section{Introduction}
\label{intro}

Dynamical billiards are an indispensable model system in the field of Hamiltonian mechanics. The amenability of billiard systems to analytical and numerical study has allowed for detailed analyses of chaotic dynamics \cite{sinai1970,bunimovich1979}, diffusion and particle transport \cite{boldrighinietal1983,moranetal1987,chernovetal1993,chernovetal2013}, the semiclassical limit \cite{grafetal1992,tomsovicheller1993,zelditchzworski1996}, and energy absorption and dissipation \cite{ulam1961,jarzynski1993,barnettetal2001,karlisetal2006,gelfreichturaev2008,karlisetal2012,dettmannleonel2013,batistic2014,demersjarzynski2015}. This last topic, the problem of energy absorption in driven billiards, was first explored by Enrico Fermi to explain the acceleration of cosmic rays \cite{fermi1949}. Since then, this ``Fermi acceleration'' and related mechanisms have been studied in contexts such as nuclear dissipation \cite{blockietal1978}, plasma physics and astrophysics \cite{kobayakawaetal2002,veltricarbone2004,biankontar2013}, and atomic optics \cite{saifetal1998}.

In this paper, we investigate energy absorption in chaotic, ergodic billiard systems, subject to a rapidly varying, time-periodic force. The system of interest is defined in Section \ref{setup}. In Section \ref{diffusion}, we argue that the evolution of the billiard particle's energy will be a diffusive process in energy space. It follows that the probability distribution for the particle energy obeys a Fokker-Planck equation in energy space, with drift and diffusion coefficients that characterize the rate at which this distribution shifts and spreads. In Section \ref{g1g2} we obtain expressions for these rates, which are found to scale like $\omega^{-2}$ for large $\omega$. In Section \ref{numerical}, we present exact (up to machine precision) numerical results that demonstrate the validity of the Fokker-Planck equation in the rapid driving regime, for the special case where the driving force is independent of position. Finally, we offer concluding remarks in Section \ref{discussion}. 

 Our results constitute a detailed case study of the process of \textit{Floquet prethermalization}, in which a periodically driven system relaxes to a thermal state with respect to an effective Hamiltonian at short to intermediate times, before ultimately gaining energy on long timescales \cite{elseetal2017,abaninetal2017,herrmannetal2017,mori2018,morietal2018,mallayyaetal2019,howelletal2019,machadoetal2019,rajaketal2019,machadoetal2020,rubio-abadal2020,pengetal2021,hodsonjarzynski2021}.  Floquet prethermalization has been widely studied as a mechanism for engineering stable, long-lived steady states of both classical and quantum systems. Within the energy diffusion framework, we obtain a comprehensive, quantitative picture of how prethermalization and its breakdown emerge from the Hamiltonian dynamics of a chaotic billiard particle.  With these results, billiard systems emerge as a valuable model system in the study of energy absorption, prethermalization, and related phenomena in periodically driven systems.

\section{Setup}
\label{setup}

We now define our system of interest. We consider a point particle of mass $m$, with position $\mathbf{x} \equiv \mathbf{x}_t$ and velocity $\mathbf{v} \equiv \mathbf{v}_t$, confined to the inside of a cavity or ``billiard.'' Precisely, the billiard is a bounded, connected subset of $d$-dimensional Euclidean space ($d \geq 2$), with a boundary or ``wall'' consisting of one or more $(d-1)$-dimensional surfaces. When strictly inside the billiard, the particle evolves smoothly according to Newton's laws. Whenever the particle reaches the billiard boundary, it undergoes an instantaneous elastic collision with the wall. 

Specifically, we assume that between collisions, the particle is subject to two forces. First, the particles experiences a conservative force $-\nabla U (\mathbf{x})$, generated by a static potential $U(\mathbf{x})$. Second, we apply a time-periodic driving force $\mathbf{F}(\mathbf{x}) \cos (\omega t) = - \nabla V (\mathbf{x}) \cos (\omega t)$, with period $T = 2 \pi /\omega$, where $V (\mathbf{x})$ is some potential. Therefore, the equations of motion for $\mathbf{x}$ and $\mathbf{v}$ are given by:

\begin{equation}
\label{newton}
\frac{d \mathbf{x}}{d t} = \mathbf{v}, \quad m \frac{d \mathbf{v}}{d t} = -\nabla U (\mathbf{x}) + \mathbf{F}(\mathbf{x}) \cos (\omega t).
\end{equation}

When the particle reaches the billiard boundary, an instantaneous elastic collision occurs. This collision leaves the position of the particle unchanged, but the component of the velocity perpendicular to the wall is instantly reversed. That is, the velocity of the particle is updated from $\mathbf{v}$ to $\mathbf{v}'$ according to the reflection law

\begin{equation}
\label{reflection}
\mathbf{v}' = \mathbf{v} - 2 (\mathbf{v} \cdot \hat{\mathbf{n}}(\mathbf{x})) \hat{\mathbf{n}}(\mathbf{x}) ,
\end{equation}

\noindent where $\hat{\mathbf{n}}(\mathbf{x})$ is the outward-facing unit vector normal to the billiard boundary at $\mathbf{x}$, the point of collision.

The equations \eqref{newton} and \eqref{reflection} fully define the dynamics of the driven particle.  We note here that our use of the term ``billiard'' is more general than the typical usage: The word ``billiard'' often simply refers to a free particle in a cavity, corresponding to the case of vanishing $U(\mathbf{x})$ and $\mathbf{F}(\mathbf{x})$. In light of this, we will use the term ``standard billiard'' to refer to the special case of $U(\mathbf{x})=0$. For a driven standard billiard, the associated \textit{undriven} billiard (obtained by additionally setting $\mathbf{F} (\mathbf{x}) = \mathbf{0}$) corresponds to a billiard in the more common sense of the word.

%The equations \eqref{newton} and \eqref{reflection} fully define the dynamics of the driven particle. We comment here that our system of interest is more general than a ``billiard'' in the typical sense of the word: The term ``billiard'' often simply refers to a free particle in a cavity, corresponding to the case of vanishing $U(\mathbf{x})$ and $\mathbf{F}(\mathbf{x})$. As described in Section \ref{discussion}, this generality allows us to apply our results to systems with many interacting particles, by exploiting the correspondence between $N$ particles in a $d$-dimensional billiard and a single particle in a $(d \times N)$-dimensional billiard. To avoid confusion stemming from this nonstandard terminology, we will use the term ``standard billiard'' to refer to the special case of $U(\mathbf{x})=0$. For a driven standard billiard, the associated \textit{undriven} billiard (obtained by additionally setting $\mathbf{F} (\mathbf{x}) = \mathbf{0}$) corresponds to a billiard in the more common sense of the word.

In our analysis, we are most interested in the evolution of the particle's energy, defined as $\mathcal{E} \equiv \mathcal{E}(\mathbf{x},\mathbf{v}) \equiv \frac{1}{2} m |\mathbf{v}|^2 + U(\mathbf{x})$. In the absence of driving, $\mathcal{E}$ is a constant of the motion: $\mathcal{E}$ is conserved under the equations of motion \eqref{newton} for $\mathbf{F} (\mathbf{x}) = \mathbf{0}$, and is also unchanged under the reflection law \eqref{reflection}. For nonzero $\mathbf{F} (\mathbf{x})$, the collisions are still energy-conserving, but \eqref{newton} implies that the particle's energy between collisions changes according to:

\begin{equation}
\label{power}
\frac{d \mathcal{E}}{d t} =  \mathbf{F}(\mathbf{x}) \cdot \mathbf{v} \cos (\omega t).
\end{equation}

\noindent In particular, we will consider the energy dynamics for large $\omega$, in the rapid driving regime.

So far, we have considered a single trajectory of the particle in the billiard. However, in our analysis, it will also be useful to consider a statistical ensemble of particles, and averages over that ensemble. Each particle trajectory in such an ensemble is determined by an initial condition $(\mathbf{x}_0,\mathbf{v}_0)$ at $t=0$, which is sampled according to some probability distribution $\rho_0(\mathbf{x}_0,\mathbf{v}_0)$ on phase space (that is, the $2d$-dimensional space of particle positions and velocities). The ensemble is then evolved in time by evolving each initial condition according to \eqref{newton} and \eqref{reflection}, yielding $\mathbf{x}_t$ and $\mathbf{v}_t$.

Any statistical property of this ensemble may be computed as an appropriate average over initial conditions, with respect to the distribution $\rho_0(\mathbf{x}_0,\mathbf{v}_0)$. In particular, since we are interested in the evolution of the system's energy $\mathcal{E}$, our analysis will focus on $\eta \equiv \eta (E,t)$, the time-dependent probability distribution for the energy. For small $dE$, $ \eta (E,t) dE$ gives the fraction of particles in the ensemble at time $t$ with energy between $E$ and $E + dE$. We may express $\eta$ as

\begin{equation}
\label{eta}
\eta (E,t) = \int d^d \mathbf{x}_0 d^d \mathbf{v}_0 \, \rho_0(\mathbf{x}_0,\mathbf{v}_0) \, \delta \Big( \mathcal{E}(\mathbf{x}_t,\mathbf{v}_t) - E \Big).
\end{equation}

\noindent Here, $d^d \mathbf{x}_0 d^d \mathbf{v}_0$ is a $2d$-dimensional infinitesimal ``hyper-volume'' element in phase space, and $(\mathbf{x}_t,\mathbf{v}_t)$ is the phase space location, at time $t$, of the trajectory with initial conditions $(\mathbf{x}_0,\mathbf{v}_0)$. For this integral, and for similar integrals in this paper unless otherwise stated, the integration over $\mathbf{x}_0$ is performed over the interior of the billiard, and the integration over $\mathbf{v}_0$ runs over all $\mathbf{v}_0 \in \mathbb{R}^d$.

The last essential assumption in our analysis is that the undriven system exhibits chaotic and ergodic motion at each energy $E$. For certain classes of undriven billiards, it has been rigorously proven that the particle motion is chaotic and ergodic \cite{sinai1970,bunimovich1979,wojtkowski1986,donnay1991}.  Although these results were derived for undriven \textit{standard} billiards, in some cases they may be extended, at least approximately, to the $U(\mathbf{x}) \neq 0$ case, e.g. by considering weak forces \cite{chernov2001}, or by invoking the correspondence between motion in a potential and free motion in non-Euclidean space \cite{beletsky1999}.

Finally, we note that if the driving force is generated by a more general time-periodic potential $V(\mathbf{x},t)$, then it is straightforward to extend our analysis by decomposing this potential as a Fourier series with fundamental frequency $\omega$. However, in order to keep the calculations relatively simple, we restrict our attention to the monochromatic driving force $\mathbf{F}(\mathbf{x}) \cos (\omega t)$.

\section{Energy diffusion}
\label{diffusion}

We now describe the evolution of the particle's energy $E$, in the limit of large $\omega$. We argue that the energy of the particle evolves diffusively in this limit. A more general and detailed version of this argument may be found in our previous paper \cite{hodsonjarzynski2021}, wherein it is shown that a \textit{generic} chaotic, ergodic Hamiltonian system will exhibit energy diffusion when subject to rapid periodic driving. Energy diffusion in chaotic billiards under rapid periodic driving is a special case of this result.

We first note that, for sufficiently large $\omega$, the effect of the driving force on the particle between collisions nearly averages to zero over a single period. This averaging effect may be rigorously demonstrated using tools such as multi-scale perturbation theory \cite{murdock1999,rahavetal2003}. However, it is also intuitively reasonable: For a very short driving period, the particle's position and velocity will remain nearly constant over the period (as long as a collision with the wall does not occur), because of the particle's inertia and finite speed. Under this approximation, integrating \eqref{newton} over a period reveals that the resulting change in position and velocity are the same as if the system was not being driven, since the term $\mathbf{F}(\mathbf{x}) \cos (\omega t)$ integrates to zero. This approximation will become better and better for shorter and shorter periods, so as $\omega$ goes to infinity, the driven evolution of the system will become closer and closer to the undriven dynamics. Notably, this conclusion holds regardless of the magnitude of $\mathbf{F} (\mathbf{x})$.

So for sufficiently large $\omega$, the drive acts as a small perturbation on the undriven dynamics. Let us choose $\omega$ large enough such that driven and undriven trajectories closely resemble one another on timescales of order $\tau_C$, the characteristic correlation time set by the undriven particle's chaotic dynamics. To show that the driven particle's energy evolves diffusively, we now consider the evolution of the particle at discrete times $t=0$, $\delta t$, $2\delta t$ ..., for some $\delta t \geq \tau_C$. Over each timestep, the particle's energy changes by a small amount $\delta E_i$, $i=1,2,3...$. Since $\delta t \geq \tau_C$, these individual energy increments will be approximately uncorrelated. That is, the particle performs a random walk along the energy axis, where each energy increment $\delta E_i$ is statistically independent of the others. On timescales much longer than $\tau_C$, after many ``steps'' in this process have occurred, such a random walk will be well-described as a process of diffusion in energy space.

If the energy of the driven particle evolves diffusively, then the particle's energy probability distribution $\eta$ (defined in \eqref{eta}) will evolve according to a Fokker-Planck equation in energy space \cite{gardiner1985}:

\begin{equation}
\label{fp}
 \frac{\partial \eta}{\partial t} = -\frac{\partial}{\partial E} \left( g_1 \eta \right)+ \frac{1}{2} \frac{\partial ^2}{\partial E^2} \left( g_2 \eta \right).
\end{equation}

\noindent Here, the drift coefficient $g_1 \equiv g_1 (E,\omega)$ and the diffusion coefficient $g_2 \equiv g_2 (E,\omega)$ characterize the diffusive process: $g_1$ gives the rate at which $\eta$ shifts along the energy axis, and $g_2$ gives the rate of diffusive spreading in energy space. In the next section, we obtain explicit expressions for $g_1$ and $g_2$ in the high frequency driving limit.  As we will see, these drift and diffusion rates are suppressed by a factor of $\omega^{-2}$ for large $\omega$.  Moreover, we show that $g_1$ is always nonnegative, which then implies that the system undergoes \textit{Fermi acceleration} \cite{fermi1949,ulam1961,barnettetal2001,karlisetal2006,gelfreichturaev2008,karlisetal2012,batistic2014} on average: The driven dynamics exhibit a statistical bias towards gaining energy, and the mean energy of the ensemble never decreases.

%We reemphasize that the energy diffusion description (and therefore the Fokker-Planck equation \eqref{fp}) is only valid on timescales much longer than the characteristic correlation time $\tau_C$ for the undriven system's chaotic evolution. For free particles in a billiard ($U(\mathbf{x}) = 0$) at speed $v = (2 E/m)^{1/2}$, $\tau_C$ is proportional to the mean free time $\lambda /v$ between collisions, where $\lambda$ is the billiard's characteristic mean free path length. Roughly speaking, correlations between states of the particle at different times will become negligible after sufficiently many collisions with the billiard wall. For a nonzero potential $U(\mathbf{x})$, the situation becomes more complex, as the decay of correlations will also depend on how rapidly the potential causes nearby trajectories to diverge from one another. 

How large must $\omega$ be for the energy diffusion picture, and the associated Fokker-Planck equation, to be approximately valid? Recall that the driving force must act as a small perturbation on the undriven dynamics. This suggests two conditions.  First, we assume that over the course of a single period, the forces $- \nabla U (\mathbf{x})$ and $\mathbf{F} (\mathbf{x}) \cos (\omega t)$ produce a very small change in the particle's velocity. If the typical magnitude of these forces is denoted by $F$, then from \eqref{newton} we can estimate that the velocity will change by an amount of order $ F/ (m \omega)$ during a period (provided a collision does not occur). We assume that this change is much smaller than $v$, the typical speed of the particle:

\begin{equation}
\label{condition1resub}
\frac{F }{m \omega}  \ll v.
\end{equation}

This is our first condition. Importantly, this ensures that when a collision occurs, the outgoing trajectory of the particle will only be slightly altered relative to the undriven motion. If \eqref{condition1resub} is not satisfied, then the particle's direction of motion will oscillate wildly back and forth due to the force $\mathbf{F}(\mathbf{x}) \cos (\omega t)$. As a result, the drive may cause the particle to collide with the wall at a substantially different angle relative to a corresponding undriven particle. The associated driven and undriven trajectories would then rapidly diverge, contrary to our requirement that the drive act as a small perturbation.

Second, we assume that the distance travelled by the particle over a typical period is very small, much smaller than any other relevant length scale associated with the system. Since \eqref{condition1resub} ensures that the particle's velocity changes little during a period, this distance travelled will be of order $ v T \sim v/\omega$. So we may write our second condition as

\begin{equation}
\label{condition2resub}
 \frac{v}{\omega} \ll l ,
\end{equation}

\noindent where $l$ is the shortest length scale in the system. $l$ may be the mean free path for the particle, or a length scale characterizing the roughness of the billiard wall, or the typical distance over which the forces $-\nabla U (\mathbf{x})$ and $\mathbf{F} (\mathbf{x})$ vary by a significant amount.

With the condition \eqref{condition2resub} satisfied, a large number of periods will occur between successive collisions with the billiard wall. Moreover, over a single period, the quantity $\mathbf{F} (\mathbf{x})$ will be nearly constant, since the particle will hardly move during this short time interval. As a result, during any period without a collision (the great majority of periods), integrating \eqref{newton} reveals that the driving force perturbs the particle's velocity by an amount $\approx \mathbf{F}(\mathbf{x}) \sin (\omega t)/(m \omega)$, and its position by $\approx -\mathbf{F}(\mathbf{x}) \left[\cos (\omega t) - 1 \right]/(m \omega^2)$. Thus, the cumulative effect of the force essentially integrates to zero as $\omega\rightarrow\infty$. Taken together, we see that if the conditions \eqref{condition1resub} and \eqref{condition2resub} are satisfied, the drive acts as a weak perturbation during periods both with and without collisions. When subject to such a drive, the particle will typically experience several collisions with the wall before its trajectory is significantly altered relative to the undriven motion.

 For any given energy $E$, which determines the typical particle speed $v$, the conditions \eqref{condition1resub} and  \eqref{condition2resub} may always be satisfied for sufficiently large $\omega$. Thus, in this rapid driving regime, we expect that the energy diffusion description will be valid over a certain range of particle energies $[ E_{min},E_{max}]$ for which these conditions hold. For a given $\omega$, the energy distribution $\eta$ for a statistical ensemble with particle energies in $[ E_{min},E_{max}]$ will evolve according to the Fokker-Planck equation \eqref{fp}. Of course, under the Fokker-Planck dynamics, this distribution will shift and spread in energy space, ultimately spreading outside of the interval $[ E_{min},E_{max}]$.  At this point, the conditions \eqref{condition1resub} and \eqref{condition2resub} are not satisfied for all particles in the ensemble.  In particular, we expect condition \eqref{condition2resub} to generally break down for sufficiently high energy particles, which are fast enough to travel a significant distance over a single period.

What happens in this high energy regime, when particle speeds have increased so that $v/\omega \sim l$? As before, condition \eqref{condition1resub} (which remains valid at high energies) tells us that the forces $- \nabla U(\mathbf{x})$ and $\mathbf{F} (\mathbf{x}) \cos (\omega t)$ only weakly perturb the particle's velocity over a given period. However, the increased speed of the particle now means that the particle travels a distance of order $v/\omega \sim l$ during this period. Assuming for simplicity that $l$ is comparable to the particle's mean free path, we see that the velocity is only slightly altered between successive collisions: Many collisions with the wall must occur before the drive significantly perturbs the particle's velocity relative to the undriven motion. Similarly, the drive will only weakly affect the particle's position: We can estimate from \eqref{newton} that the drive will perturb the particle's position by an amount of order $F/(m \omega^2)$ during a period, which by \eqref{condition1resub} and $v/\omega \sim l$ is much smaller than $l$. Therefore, the energy diffusion description may still be valid at energies greater than $E_{max}$, since we can potentially treat the drive as a small perturbation on the undriven dynamics. With that said, our main focus in this paper is rapidly driven particles, for which the conditions \eqref{condition1resub} and \eqref{condition2resub} are \textit{both} satisfied. In particular, the expressions for $g_1$ and $g_2$ obtained in the next section are only valid in this regime.

 This process of energy diffusion may be understood in terms of the phenomenon of Floquet prethermalization \cite{elseetal2017,abaninetal2017,herrmannetal2017,mori2018,morietal2018,mallayyaetal2019,howelletal2019,machadoetal2019,rajaketal2019,machadoetal2020,rubio-abadal2020,pengetal2021,hodsonjarzynski2021}.  Consider an ensemble of driven particles with initial energy $E_0$, for which the conditions \eqref{condition1resub} and \eqref{condition2resub} are satisfied. The energy evolution of this ensemble may be divided into three stages. First, since the system is only weakly perturbed by the drive, the particles in the ensemble will exhibit chaotic, ergodic motion at nearly constant energy $E_0$. These dynamics lead to a process of chaotic mixing, in which the ensemble is distributed microcanonically (see \eqref{micro}) over a surface of constant energy in phase space \cite{dorfman1999}. That is, the system thermalizes at energy $E_0$: This is the prethermal stage. Second, the particle's energy distribution $\eta$ will slowly shift and broaden, as energy diffusion occurs according to the Fokker-Planck equation \eqref{fp}. As a result of this energy spreading, conditions \eqref{condition1resub} and \eqref{condition2resub} will eventually not hold for a significant fraction of particles in the ensemble, as $\eta$ spreads outside the interval $[ E_{min},E_{max}]$.  At this point, although the energy evolution may still be diffusive, the energy drift and diffusion rates will no longer be given by the expressions \eqref{g1} and \eqref{g2} in the next section. In this third and final stage, general plausibility arguments for the existence of Fermi acceleration in driven billiards (see, e.g., Fermi's original work \cite{fermi1949}) lead us to speculate that on-average energy growth will continue, at least for certain choices of billiards and driving forces. In particular, since the particle's displacement during a period can be comparable to the typical distance travelled between collisions, resonances between the particle motion and the drive may result in especially rapid energy growth.  Because the phase space of a billiard particle is unbounded, this energy absorption may potentially continue without limit.

\section{Drift and diffusion coefficients}
\label{g1g2}

We now derive expressions for the drift and diffusion coefficients $g_1$ and $g_2$, in the limit of large $\omega$. We calculate these quantities in terms of powers of the small parameter $\omega^{-1}$, and ultimately only retain terms of order $O(\omega^{-2})$, the lowest non-zero order. We compute $g_2$ in terms of the variance in energy acquired by an ensemble of particles, initialized in a microcanonical ensemble at $t=0$ and then subject to the rapid drive. Then, we use the fluctuation-dissipation relation \eqref{fd}, established in \cite{hodsonjarzynski2021}, which allows us to calculate $g_1$ from our knowledge of $g_2$.

To compute $g_2$, suppose that the initial conditions of the particle at $t=0$ are sampled according to a microcanonical distribution at energy $\mathcal{E} = E_0$. In the microcanonical ensemble, particles are confined to a single energy shell in phase space (a surface of constant energy), and the distribution of particles on this shell is uniform with respect to the Liouville measure. The initial distribution $\rho_0 (\mathbf{x}_0,\mathbf{v}_0) = \rho_{E_0} (\mathbf{x}_0,\mathbf{v}_0)$ corresponding to this ensemble is given by
\begin{equation}
\label{micro}
\rho_{E_0} (\mathbf{x}_0,\mathbf{v}_0) \equiv \frac{1}{\Sigma (E_0)} \delta \Big( \mathcal{E}(\mathbf{x}_0,\mathbf{v}_0) - E_0 \Big),  
\end{equation}

\begin{equation}
\label{dos}
\Sigma(E) \equiv \int d^d \mathbf{x} d^d \mathbf{v} \, \delta \Big( \mathcal{E}(\mathbf{x},\mathbf{v}) - E \Big),
\end{equation}

\noindent where $\Sigma (E)$ is the density of states for the undriven system. Since all particles in this ensemble have energy $E_0$ at $t=0$, this initial distribution corresponds to an initial condition $\eta (E,0) = \delta (E - E_0)$ for the Fokker-Planck equation \eqref{fp}.

We now allow the driven system to evolve for a time $\Delta t$, where $\Delta t$ is long enough that the energy evolution is diffusive (i.e., $\Delta t \gg \tau_C$), but short enough that the change in the particle's energy is still small. By the end of this time interval, the ensemble of particles will have acquired a variance in energy $ \mathrm{Var} (\mathcal{E}) \equiv \langle \mathcal{E}^2 \rangle - \langle \mathcal{E} \rangle^2 =\langle \left(\Delta \mathcal{E}\right)^2 \rangle - \langle \Delta \mathcal{E} \rangle^2  $, where $\langle ... \rangle$ denotes an average over the ensemble at $t = \Delta t$, and $\Delta \mathcal{E} \equiv \mathcal{E} - E_0$ is the energy change of the particle from $t=0$ to $t= \Delta t$. From the Fokker-Planck equation \eqref{fp}, given the initial condition $\eta(E,0)=\delta(E-E_0)$, it follows that \cite{gardiner1985}
\begin{equation}
\label{fpvar}
\mathrm{Var} (\mathcal{E}) \approx g_2 (E_0,\omega) \Delta t.
\end{equation}
\noindent Therefore, to determine $g_2 (E_0,\omega)$ for any particular $E_0$, it is sufficient to calculate $\mathrm{Var} (\mathcal{E})$, with trajectories sampled according to the appropriate microcanonical distribution $\rho_{E_0} (\mathbf{x}_0,\mathbf{v}_0)$.

This calculation may be summarized as follows, with details given below and in Appendix A. First, for a given trajectory in the ensemble over the time $\Delta t$, we evaluate the associated energy change $\Delta \mathcal{E}$. From the fact that the drive acts as a small perturbation for large $\omega$, it follows that the dominant contribution to $\Delta \mathcal{E}$ is associated with driving periods during which a collision occurs. These $O(\omega^{-1})$ contributions are given by \eqref{dEicollision2}. We then average over the ensemble to obtain  $\mathrm{Var} (\mathcal{E})$. Since the energy changes associated with different collisions become uncorrelated in the rapid driving limit, this average simplifies to \eqref{varE2}, as shown in Appendix A. Finally, we express this result in terms of an integral over the billiard boundary, leading to the expression \eqref{g2} for $g_2$.

To begin, let us consider $\Delta \mathcal{E}$ for a particular particle in the ensemble. We may view this energy change as a sum of the $M= \Delta t/T$ small energy changes that occur over each period of the drive (assuming, for simplicity, that $\Delta t$ is an integer multiple of the period $T$). For sufficiently small $T$, \textit{at most} one collision will occur over each driving period. This property is guaranteed for a typical trajectory by condition \eqref{condition2resub}.  Therefore, in this regime, $\Delta \mathcal{E}$ is a sum of two contributions: Energy changes from periods with no collisions, and energy changes from periods with a single collision. We will examine these two possibilities in turn.

First, suppose that no collision occurs during the $i^{\mathrm{th}}$ period, from $t=(i-1)T$ to $t = i T$, with associated energy change $\Delta \mathcal{E}_i$. If we integrate \eqref{power} over this period and perform an integration by parts, we find that the boundary terms vanish, and the resulting expression for $\Delta \mathcal{E}_i$ is:
\begin{equation}
\begin{split}
\label{dEinocollision}
\Delta \mathcal{E}_i &= -\omega^{-1} \int_{(i-1)T}^{i T} dt \, \frac{d}{d t} \left[ \mathbf{F} (\mathbf{x}_t) \cdot \mathbf{v}_t \right] \sin (\omega t) \\ &=  -\omega^{-1} \int_{(i-1)T}^{i T} dt \,   \Bigg[ \mathbf{v}_t \cdot D\mathbf{F} (\mathbf{x}_t) \mathbf{v}_t \Bigg. \\ &\quad \Bigg. - \frac{\nabla U (\mathbf{x}_t) \cdot \mathbf{F}(\mathbf{x}_t)}{m} + \frac{|\mathbf{F}(\mathbf{x}_t)|^2}{m} \cos (\omega t) \Bigg]  \sin (\omega t) .
\end{split}
\end{equation}
In moving from the first line to the second line, we have used the equations of motion \eqref{newton} to evaluate the derivative $d \left[ \mathbf{F} (\mathbf{x}_t) \cdot \mathbf{v}_t \right] /dt$. The symbol $D \mathbf{F} (\mathbf{x})$ denotes the Jacobian matrix for the function $\mathbf{F} (\mathbf{x})$, with matrix elements $[D\mathbf{F} (\mathbf{x})]_{ij} \equiv \partial F_i / \partial x_j$, where $x_i$ and $F_i$ are the $i^{\mathrm{th}}$ components of $\mathbf{x}$ and $\mathbf{F} (\mathbf{x})$.

So far, this is exact. Let us estimate the size of this quantity, in terms of orders of the small parameter $\omega^{-1}$. Since there is a factor of $\omega^{-1}$ outside the second integral in \eqref{dEinocollision}, and since we are integrating over a single period of duration $T = O(\omega^{-1})$, $\Delta \mathcal{E}_i$ is at most an $O(\omega^{-2})$ quantity. To approximate $\Delta \mathcal{E}_i$, we may replace $\mathbf{x}_t$ and $\mathbf{v}_t$ in the integrand by their values at the beginning of the period. Since $\mathbf{x}_t$ and $\mathbf{v}_t$ change over a period by an amount of order $O(\omega^{-1})$, the resulting expression for $\Delta \mathcal{E}_i$ is valid up to corrections of order $O(\omega^{-3})$. After this replacement, we are simply integrating the functions $\sin (\omega t)$ and $\cos (\omega t)\sin (\omega t)$ over a single period, which both vanish. Thus, $\Delta \mathcal{E}_i$ is a $O(\omega^{-3})$ quantity. Of course, the number of periods in which no collision occurs will scale like $\omega$; thus, the total energy change associated with collisionless periods is of order $O(\omega^{-2})$.

The periods during which a collision takes place are more interesting. Suppose that the particle experiences $N$ collisions between $t=0$ and $t=\Delta t$, at times $t_1, t_2 ... t_N$. If the $k^{\mathrm{th}}$ collision occurs during the $i^{\mathrm{th}}$ period, then integrating \eqref{power} over this period yields the associated energy change $\Delta \mathcal{E}_i$:
\begin{equation}
\begin{split}
\label{dEicollision}
\Delta \mathcal{E}_i &=  \int_{(i-1)T}^{t_k} dt \,  \mathbf{F} (\mathbf{x}_t) \cdot \mathbf{v}_t \cos (\omega t) \\ &\quad +  \int_{t_k}^{i T} dt \,  \mathbf{F} (\mathbf{x}_t) \cdot \mathbf{v}_t \cos (\omega t) .
\end{split}
\end{equation}
Each integral above is over a fraction of the period, and is therefore of order $O(\omega^{-1})$. By the same logic that we used for the collisionless case, we may approximate $\mathbf{F}(\mathbf{x}_t)$ and $\mathbf{v}_t$ in the first integral by $\mathbf{F}_k$ and $\mathbf{v}_k$, their values instantaneously prior to the $k^{\mathrm{th}}$ collision. Similarly, $\mathbf{F}(\mathbf{x}_t)$ and $\mathbf{v}_t$ in the second integral can be approximated by $\mathbf{F}_k$ and $\mathbf{v}_k^+$, where $\mathbf{v}_k^+$ is the particle's velocity immediately after the collision. The reflection law \eqref{reflection} tells us that $\mathbf{v}_k^+ = \mathbf{v}_k - 2 \left( \mathbf{v}_k \cdot \hat{\mathbf{n}}_k \right) \hat{\mathbf{n}}_k$, where $\hat{\mathbf{n}}_k$ is the normal to the wall at the point of collision.

Upon making these substitutions, the resulting approximation for $\Delta \mathcal{E}_i$ is valid up to corrections of order $O(\omega^{-2})$. The integrals over $\cos (\omega t)$ are easily evaluated, and we obtain: 

\begin{equation}
\label{dEicollision2}
\Delta \mathcal{E}_i =  2 \omega^{-1}\left( \mathbf{F}_k \cdot \hat{\mathbf{n}}_k \right) \left( \mathbf{v}_k \cdot \hat{\mathbf{n}}_k \right) \sin (\omega t_k)  + O(\omega^{-2}).
\end{equation}

\noindent Therefore, each collision that occurs is accompanied by a corresponding energy change of order $O(\omega^{-1})$ over the associated period, given by the above expression. Moreover, since the total energy change associated with \textit{collisionless} periods is of order $O(\omega^{-2})$, the energy changes corresponding to collisions are the dominant contribution to $\Delta \mathcal{E}$ for large $\omega$. After summing over all $N$ collisions to obtain $\Delta \mathcal{E}$, we can substitute this result into $ \mathrm{Var} (\mathcal{E}) = \langle \left(\Delta \mathcal{E}\right)^2 \rangle - \langle \Delta \mathcal{E} \rangle^2  $:

\begin{widetext}
\begin{equation}
\label{varE0}
\mathrm{Var} (\mathcal{E}) = 4 \omega^{-2} \Biggl< \left[\sum_{k=1}^N \left( \mathbf{F}_k \cdot \hat{\mathbf{n}}_k \right) \left( \mathbf{v}_k \cdot \hat{\mathbf{n}}_k \right) \sin (\omega t_k) \right]^2 \Biggr>  - 4 \omega^{-2} \Biggl< \sum_{k=1}^N \left( \mathbf{F}_k \cdot \hat{\mathbf{n}}_k \right) \left( \mathbf{v}_k \cdot \hat{\mathbf{n}}_k \right) \sin (\omega t_k) \Biggr>^2 + O(\omega^{-3}).
\end{equation}
\end{widetext}

This expression is computed in Appendix A. In this calculation, we find that the oscillating factors $\sin (\omega t_k)$ are uncorrelated with one another, and with the quantities $\left( \mathbf{F}_k \cdot \hat{\mathbf{n}}_k \right) \left( \mathbf{v}_k \cdot \hat{\mathbf{n}}_k \right)$, for large $\omega$. The phases $\omega t_k \, \mathrm{mod} \, 2 \pi$ may be thought of as effectively independent random variables, uniformly distributed on $[0,2 \pi )$. Intuitively, this lack of correlation arises because otherwise similar trajectories in the ensemble may have totally different values of $\sin (\omega t_k)$: Two nearby trajectories with even a small difference between the associated collision times $t_k$ will have a huge $O(\omega)$ difference in the value of $\omega t_k$, for large $\omega$.

As a result, averages over the oscillating factors $\sin(\omega t_k)$ are found to vanish. The only non-vanishing terms in \eqref{varE0} are the ``diagonal'' terms in $ \langle \left(\Delta \mathcal{E}\right)^2 \rangle  $, which include a factor of $\sin^2 (\omega t_k)$ that averages to $1/2$. We are left with:

\begin{equation}
\label{varE2}
\mathrm{Var} (\mathcal{E}) = 2 \omega^{-2} \Biggl<  \sum_{k=1}^N   \left( \mathbf{F}_k \cdot \hat{\mathbf{n}}_k \right)^2 \left( \mathbf{v}_k \cdot \hat{\mathbf{n}}_k \right)^2 \Biggr>_0 + O(\omega^{-3}).
\end{equation}

\noindent Here, the subscript $0$ denotes that the average is now taken over an ensemble of \textit{undriven} trajectories, evolved with $\mathbf{F} (\mathbf{x}) = \mathbf{0}$. The error accrued by replacing the true driven trajectories with their undriven counterparts is of order $O(\omega^{-3})$, so we neglect it.

Then, using standard techniques for evaluating ensemble averages in billiard systems, we may express this average as an integral over the billiard boundary. We simply present the results here; the details of this calculation are also found in Appendix A. Let $d S $ denote an infinitesimal $(d-1)$-dimensional patch of ``surface'' or ``hyper-area'' of the billiard wall, surrounding a location $\mathbf{x}$ on the wall. Such a patch has an associated outward-facing normal vector $\hat{\mathbf{n}} \equiv \hat{\mathbf{n}} (\mathbf{x})$, defined as in \eqref{reflection}, and an associated value of $\mathbf{F} \equiv \mathbf{F} (\mathbf{x})$. We may express $\mathrm{Var} (\mathcal{E})$ as an integral over all such patches:

\begin{equation}
\label{varE3}
 \mathrm{Var} (\mathcal{E}) = \frac{4 \omega^{-2} \Delta t}{d+1}  \int dS  \, \gamma_{E_0}  v_{E_0}^2 \left( \mathbf{F} \cdot \hat{\mathbf{n}}\right)^2 + O (\omega^{-3}).
\end{equation}
Here, we define $v_E \equiv v_E(\mathbf{x})$ as

\begin{equation}
\label{defvE}
v_E(\mathbf{x}) \equiv 
\begin{cases}
\left[ 2 \left( E - U(\mathbf{x})\right)/m \right]^{1/2} & \text{if $U(\mathbf{x}) \leq E$} \\
0 & \text{otherwise}
\end{cases}
\end{equation}
which for $U(\mathbf{x}) \leq E$ is the speed of an undriven particle at position $\mathbf{x}$ with energy $E$. $\gamma_E \equiv \gamma_E (\mathbf{x})$ is the average collision rate per unit hyper-area of the billiard boundary for particles at position $\mathbf{x}$, averaged over undriven particles in the microcanonical ensemble at energy $E$. As explained in Appendix A, an explicit expression for $\gamma_E (\mathbf{x}) $ is given by
\begin{equation}
\label{gamma}
\gamma_E (\mathbf{x}) = \frac{B_{d-1}}{m} \frac{v_E (\mathbf{x})^{d-1}}{\Sigma (E)},
\end{equation}
\noindent where $B_n = \pi^{n/2} / \Gamma \left(\frac{n}{2} + 1 \right)$ is the hyper-volume of the unit ball in $n$-dimensional space, and where $\Sigma (E)$ is the density of states defined in \eqref{dos}.  $\Gamma (s)$ is the gamma function, which coincides with the factorial $(s-1)!$ for positive integers $s$.

Upon comparing \eqref{varE3} with \eqref{fpvar}, and relabelling $E_0$ as $E$, we obtain our final expression for $g_2$:
\begin{equation}
\label{g2}
 g_2 (E,\omega) =  \frac{4 \omega^{-2}}{d+1} \,   \int dS  \, \gamma_E  v_E^2 \left( \mathbf{F} \cdot \hat{\mathbf{n}} \right)^2  .
\end{equation}
In this equation and in the remainder of this section, we suppress the $O(\omega^{-3})$ corrections. Notably, the above expression can be computed without any knowledge of the particle trajectories, and depends on $\mathbf{F} (\mathbf{x}) $ only via the value of this force at the boundary of the billiard. This special dependence on $\mathbf{F} (\mathbf{x}) $ is sensible, since we know that the dominant changes in the particle's energy are associated with collisions with the wall.  Also, we emphasize that while the potential $U(\mathbf{x})$ does not appear explicitly in \eqref{g2}, $g_2$ does depend on $U(\mathbf{x})$ via the quantities $v_E (\mathbf{x})$ and $\gamma_E (\mathbf{x})$.

To calculate the drift coefficient $g_1$, we use the following fluctuation-dissipation relation derived in \cite{hodsonjarzynski2021} for general chaotic Hamiltonian systems:
\begin{equation}
\label{fd}
g_1(E,\omega) = \frac{1}{2 \Sigma (E)} \frac{\partial}{\partial E} \Big[ g_2(E,\omega) \Sigma(E) \Big].
\end{equation}
This relation emerges as a consequence of Liouville's theorem.
Substituting \eqref{defvE} - \eqref{g2} into \eqref{fd}, we arrive at our final expression for the drift coefficient:

\begin{equation}
\label{g1}
 g_1 (E,\omega) =  \frac{2 \omega^{-2}}{m}   \int dS  \, \gamma_E  \left( \mathbf{F}  \cdot \hat{\mathbf{n}} \right)^2  .
\end{equation}

\noindent This result implies that $g_1$ is always nonnegative (up to the $O(\omega^{-3})$ corrections), since $\gamma_E (\mathbf{x}) \geq 0$ for all $\mathbf{x}$ on the billiard boundary. From the Fokker-Planck equation \eqref{fp}, it follows that $d \langle \mathcal{E}  \rangle / d t = \langle g_1 (\mathcal{E},\omega) \rangle$, where the ensemble average $\langle ... \rangle$ is given by $\langle f(\mathcal{E}) \rangle = \int dE \, \eta (E,t) f(E)$ for any function $f(\mathcal{E})$. Therefore, \eqref{g1} implies that the average energy of particles in the ensemble never decreases; that is, the system exhibits \textit{Fermi acceleration} on average.

Combined with the expressions \eqref{g1} and \eqref{g2} for $g_1$ and $g_2$, the Fokker-Planck equation \eqref{fp} now fully characterizes the diffusive dynamics of the particle's energy under high frequency driving.  Note that these expressions are only valid for energies in the range $[ E_{min}, E_{max}]$, for which conditions \eqref{condition1resub} and \eqref{condition2resub} both hold. For energies above $E_{max}$, the condition \eqref{condition2resub} breaks down, and the $O(\omega^{-3})$ corrections can no longer be ignored.  Also, as mentioned previously, all of the above arguments and calculations may also be generalized to a broader class of periodic driving forces.

In the remainder of this section, we set $U(\mathbf{x})=0$ in order to evaluate $g_1$ and $g_2$ for a standard billiard. In this case, the undriven particle maintains a constant speed $v_E=\sqrt{2E/m}$, independent of position. This greatly simplifies the calculation of both the density of states $\Sigma(E)$ and the collision rate $\gamma_E (\mathbf{x})$ -- note that \eqref{micro} factorizes into two $d$-dimensional integrals, over position and velocity.
We obtain
\begin{equation}
    \Sigma (E) = d B_d  \dfrac{V v_E^{d-2}}{m},\quad
    \gamma_E  = \dfrac{1}{d} \dfrac{B_{d-1}}{B_d} \dfrac{v_E}{V},
\end{equation}
where $V$ is the $d$-dimensional hyper-volume of space enclosed by the billiard.
Our expressions for the drift and diffusion coefficients now become
\begin{align}
\label{g1noU}
   g_1 (E,\omega)  &=  \frac{2 \omega^{-2} v_E}{m \lambda} \,   \frac{1}{S} \int dS  \,  \left( \mathbf{F}  \cdot \hat{\mathbf{n}} \right)^2 \qquad ( U = 0) \\
   \label{g2noU}
   g_2 (E,\omega)  &=  \frac{4 \omega^{-2} v_E^3}{(d+1)\lambda} \,   \frac{1}{S} \int dS  \,  \left( \mathbf{F}  \cdot \hat{\mathbf{n}} \right)^2 \qquad ( U = 0)
\end{align}
where $S$ denotes the $(d-1)$-dimensional hyper-area of the billiard boundary, and $\lambda \equiv d \dfrac{B_d}{B_{d-1}} \dfrac{V}{S}$ is the mean free path (the average distance between collisions) of the undriven billiard particle \cite{chernov1997}.

In \eqref{g1noU} and \eqref{g2noU}, the dependence of $g_1$ and $g_2$ on the particle energy $E$ enters only through the quantity $v_E=\sqrt{2E/m}$.
Focusing specifically on energy absorption, we obtain, using the relation $d \langle \mathcal{E}  \rangle / d t = \langle g_1 (\mathcal{E},\omega) \rangle$,
\begin{equation}
    \frac{d\langle{\mathcal E}\rangle}{dt} =  \frac{2 \bar{v}(t)}{m\lambda \omega^2} \frac{1}{S} \int dS  \,  \left( \mathbf{F}  \cdot \hat{\mathbf{n}} \right)^2 \qquad ( U = 0)
\end{equation}
where $\bar{v}(t)\equiv \int dE\,\eta(E,t)\, v_E(E)$ is the mean particle speed at time $t$.
Thus the average rate of energy absorption is proportional to the mean particle speed and inversely proportional to the square of the driving frequency, with a constant of proportionality determined by the particle mass, the shape and dimensionality of the billiard, and the driving field $\mathbf{F}(\mathbf{x})$.
For a three-dimensional billiard this result becomes
\begin{equation}
    \frac{d\langle{\mathcal E}\rangle}{dt} =  \frac{\bar{v}(t)}{2 m \omega^2 V} \,\int dS  \,  \left( \mathbf{F}  \cdot \hat{\mathbf{n}} \right)^2 \qquad ( U = 0, d=3)
\end{equation}

This expression resembles the {\it wall formula}, a semiclassical estimate of dissipation in low-energy nuclear processes, which gives a dissipation rate proportional to mean particle speed, with a constant of proportionality that includes a surface integral over the boundary of the nucleus; see Eq. (1.2) of Ref.~\cite{blockietal1978}.
This resemblance is not surprising, since in both cases the system's energy evolves via an accumulation of small changes, sometimes positive, sometimes negative, occurring at collisions between the particle and the billiard boundary.
In fact, the wall formula can be derived within an energy diffusion approach analogous to the one developed above~\cite{jarzynski1993}.

\section{Numerical results}
\label{numerical}

\begin{figure}[ht]
\centering
\includegraphics[width=0.45\textwidth]{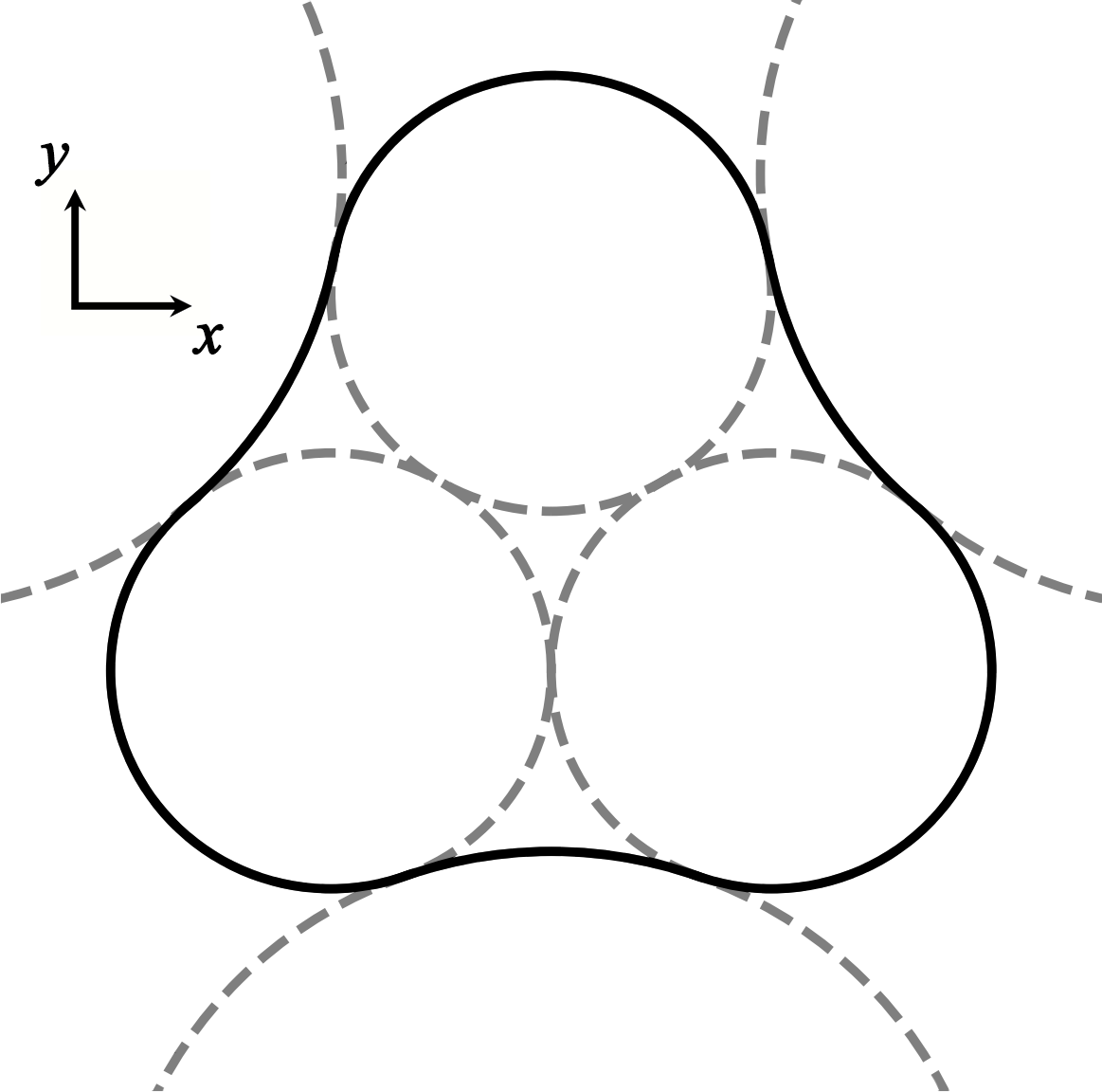}
\caption{Diagram of the clover billiard, constructed from six mutually tangent circles. The billiard boundary is given by the solid line. The inner circles have radius $R_1 = 1$, and the outer circles have radius $R_2=2.$}
\label{fig:clover}
\end{figure}

We now present numerical simulation results that corroborate our calculations. We consider the special case of a particle in a two-dimensional ``clover'' billiard, (see Figure \ref{fig:clover}), subject only to a time-periodic, spatially uniform force. Since a \textit{free} particle in the clover billiard is known \cite{jarzynski1993} to exhibit chaotic and ergodic motion, this system satisfies all the assumptions of our paper, as long as the drive is sufficiently rapid.  Specifically, in the equations of motion \eqref{newton}, we set $U(\mathbf{x}) = 0$, and take $\mathbf{F} (\mathbf{x}) = \mathbf{F} $ to be independent of position.  This special case is particularly amenable to simulation, since the motion of the particle between collisions may be computed exactly. Morever, as described in Appendix B, the Fokker-Planck equation admits an explicit analytical solution for this choice of $U(\mathbf{x})$ and $\mathbf{F}(\mathbf{x})$. 

For this system, we calculate the evolution of the energy distribution $\eta (E,t)$ in two ways: By directly evolving an ensemble of particle trajectories according to \eqref{newton} and \eqref{reflection}, and by solving the Fokker-Planck equation \eqref{fp}. If the energy diffusion description is accurate, then the energy distributions obtained in both cases will coincide. We present the results  of these computations  here, and leave the details of our calculations to Appendix B. 

To test our model, we evolve an ensemble of driven particles with mass $m=1$ in the clover billiard, with $R_1 = 1$ and $R_2 = 2$ (see Figure \ref{fig:clover}). The mean free path for particles in this billiard is $\lambda \approx 2.610$, as shown in  Appendix B. The particles are initialized at $t=0$ with speed $v_0=1$, in a microcanonical ensemble at energy $E_0 = m v_0^2 /2 = 1/2$. We set $\mathbf{F} = F(\hat{\mathbf{x}} + \hat{\mathbf{y}})/\sqrt{2}$, where $\hat{\mathbf{x}}$ and $\hat{\mathbf{y}}$ are the unit vectors for the coordinate system in Figure \ref{fig:clover}, and choose $F = |\mathbf{F}| = 10$. We run simulations for a range of driving frequencies $\omega$, with a focus on the high-frequency driving regime.

\begin{figure*}[ht]
\centering
\includegraphics[width=1.0\textwidth]{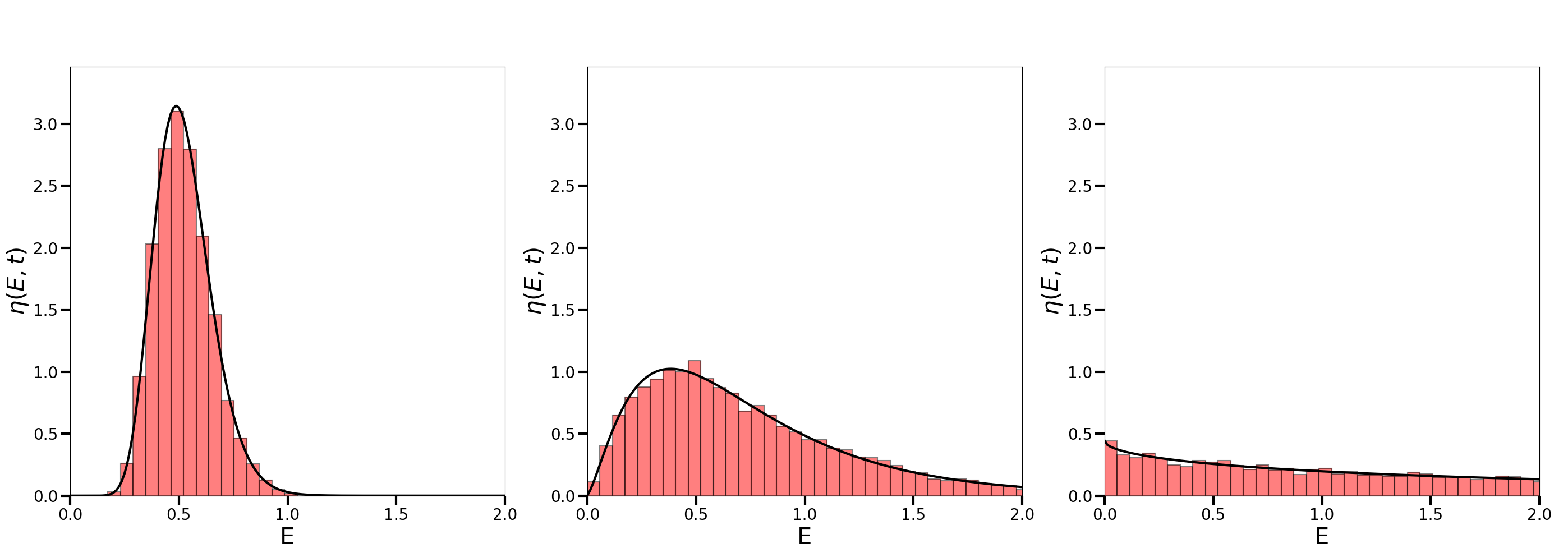}
\caption{Evolution of an ensemble starting with energy $E_0 =1/2$, with $F=10$ and $\omega = 40 \pi$. The three snapshots are captured at $t=10$, $t=100$, and $t=1000$. The histograms are populated from the numerical simulations, and the solid lines are the solution to the Fokker-Planck equation.}
\label{fig:histw40pi}
\end{figure*}

\begin{figure*}[ht]
\centering
\includegraphics[width=1.0\textwidth]{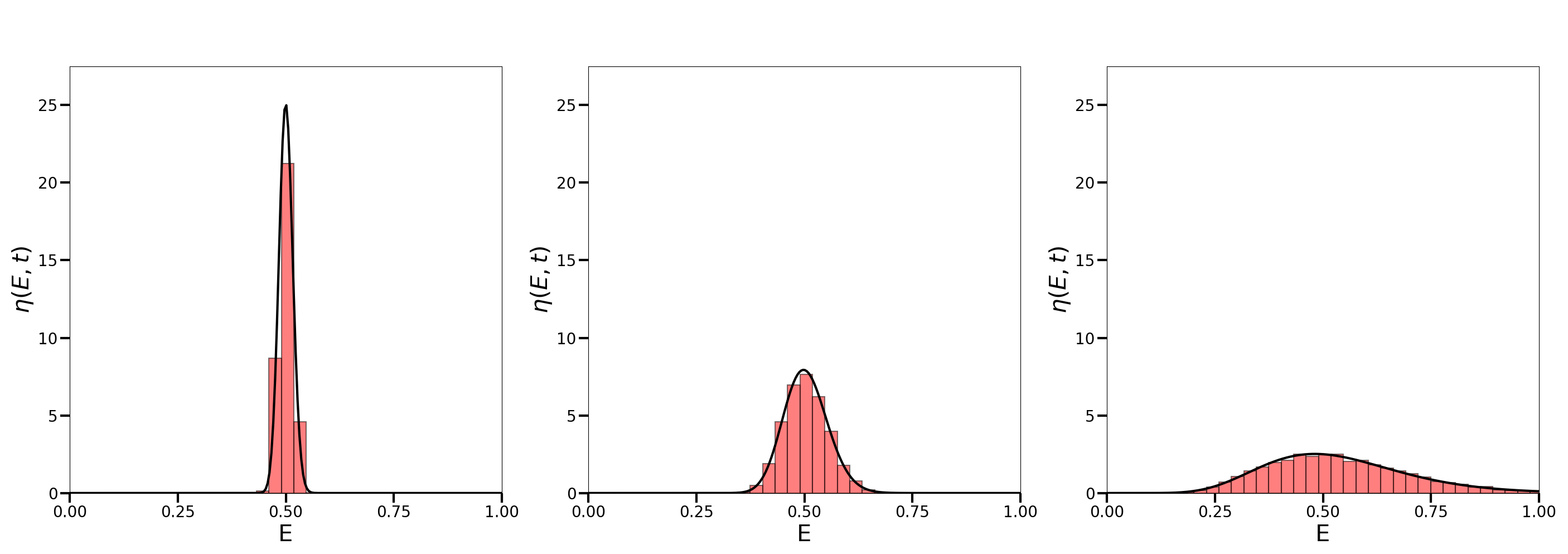}
\caption{
Same as Fig.~\ref{fig:histw40pi}, but with $\omega=320 \pi$, and with a different scaling of the axes.}
\label{fig:histw320pi}
\end{figure*}

First, we verify the validity of the Fokker-Planck equation. For various values of $\omega$, we evolve an ensemble of $N=10^5$ driven particles, and then compare the energy distribution of this ensemble with the energy distribution obtained by solving the Fokker-Planck equation. The plots in Figures \ref{fig:histw40pi} and \ref{fig:histw320pi} illustrate this comparison at times $t=10$, $100$, and $1000$, for driving frequencies $\omega = 40 \pi$ and $\omega = 320 \pi$ (note that the conditions \eqref{condition1resub} and \eqref{condition2resub}  are satisfied for these parameter choices). We find close agreement between the true energy distribution (represented by the histograms) and the Fokker-Planck energy distribution (the solid lines).

\begin{figure*}[ht]
\centering
\includegraphics[width=1.0\textwidth]{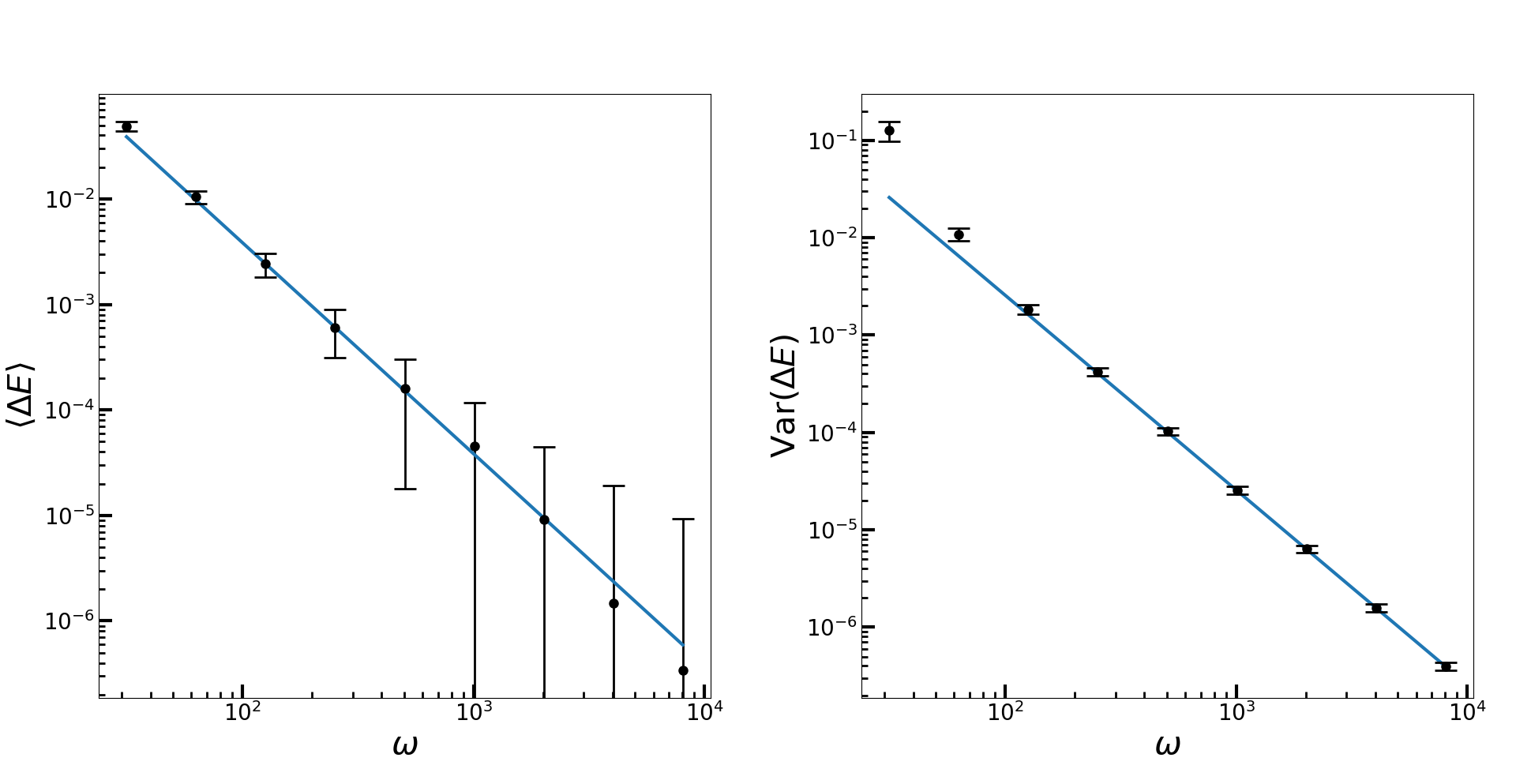}
\caption{$\langle \Delta \mathcal{E} \rangle $ and $\mathrm{Var}(\Delta \mathcal{E})$ versus $\omega$, for an initial ensemble with energy $E_0 = 1/2$, with $F = 10$ and $\Delta t = 20$. The points denote results of the numerical simulations, and the solid line corresponds to the theoretical predictions given by \eqref{g1noU} and \eqref{g2noU}.}
\label{fig:deltaEVarEvsw}
\end{figure*}

Second, we look specifically at the ensemble mean $\langle \Delta \mathcal{E}\rangle$ and variance $\mathrm{Var} (\Delta \mathcal{E})$ of the energy change $\Delta \mathcal{E}$. If a microcanonical ensemble of initial conditions at energy $E_0$ is evolved for a short time $\Delta t$, then the Fokker-Planck equation predicts that $\langle \Delta \mathcal{E}\rangle \approx g_1 (E_0, \omega) \Delta t$ and $\mathrm{Var} (\Delta \mathcal{E}) \approx g_2 (E_0, \omega) \Delta t$ \cite{gardiner1985}. Here, $\Delta t$ must be longer than the correlation timescale associated with the particle's undriven motion, but short enough that the relative change in the energy of any particle in the ensemble is still very small. To test this theoretical result, we evolve an ensemble of $N=10^6$ driven particles for a time $\Delta t = 20$, and then compute the resulting values of $\langle \Delta \mathcal{E} \rangle$ and $\mathrm{Var} (\Delta \mathcal{E})$. We repeat this for a range of driving frequencies from $\omega = 10 \pi$ to $\omega =2560 \pi$, and then plot $\langle \Delta \mathcal{E} \rangle$ and $\mathrm{Var} (\Delta \mathcal{E})$ versus $\omega$ in Figure \ref{fig:deltaEVarEvsw}. For sufficiently large $\omega$, the true values of $\langle \Delta \mathcal{E} \rangle$ and $\mathrm{Var} (\Delta \mathcal{E})$ are in good agreement with the theoretical predictions $\langle \Delta \mathcal{E}\rangle \approx g_1 (E_0, \omega) \Delta t$ and $\mathrm{Var} (\Delta \mathcal{E}) \approx g_2 (E_0, \omega) \Delta t$, where $g_1$ and $g_2$ are given by the formulas \eqref{g1noU} and \eqref{g2noU}. Note that for large $\omega$, the error bars in Figure \ref{fig:deltaEVarEvsw} associated with $\langle \Delta \mathcal{E} \rangle$ become very large. This is because the fluctuations in $\Delta \mathcal{E}$ about its average are on the order of $\sqrt{\mathrm{Var} (\Delta \mathcal{E})} = O(\omega^{-1})$, while $\langle \Delta \mathcal{E} \rangle = O(\omega^{-2})$ itself is much smaller.

We note that the value of $F = 10$ corresponds to a ``strong'' driving force, in the following sense. Suppose that we set $\omega = 0$, so that the driving force is time-\textit{independent}, and then estimate the change in a particle's energy as it moves across the billiard. In the $\omega = 0$ case, the particle simply experiences free-fall within the billiard, with a uniform gravitational field pointing in the direction of $\mathbf{F} = F(\hat{\mathbf{x}} + \hat{\mathbf{y}})/\sqrt{2}$. If we initialize the particle on one side of the billiard and let it ``fall'' to the other side, then the (kinetic) energy gained by the particle during its descent will be given by $\Delta E = F \Delta x$, where $\Delta x$ is the distance that the particle moves in the direction of $\mathbf{F}$. $\Delta x$ will be on the order of the mean free path $\lambda \approx 2.610$, and so we find $\Delta E \sim 26$. This energy change is an order of magnitude larger than the particle's initial energy $E_0 = 1/2$. Clearly, when $\omega = 0$ (or generally, if $\omega$ is small), the driving force has a very large effect on the particle trajectories, and therefore we should not expect an energy diffusion description to apply. For $F=10$, we should only expect energy diffusion for sufficiently large values of $\omega$. Testing our model with this value of $F$ thus insures that energy diffusion is really a consequence of rapid driving, and not simply the result of a weak driving force.

\section{Discussion}
\label{discussion}

We have fully characterized the diffusive evolution of energy in chaotic, ergodic billiard systems subject to a rapid periodic driving force. We obtained the associated energy drift and diffusion rates up to second order in the small parameter $\omega^{-1}$, and corroborated our theoretical predictions with numerical simulations of a driven particle in a clover-shaped billiard.  We now conclude with a discussion of connections between this paper and other work, and of possible future directions for research.

First, as described in Section \ref{diffusion}, our model is a detailed case study of \textit{Floquet prethermalization}, and its ultimate breakdown due to energy absorption. Floquet prethermalization, a phenomenon which has been documented in a range of classical and quantum systems, occurs when a periodically driven system relaxes to a long-lived thermal state with respect to an effective Hamiltonian \cite{elseetal2017,abaninetal2017,herrmannetal2017,mori2018,morietal2018,mallayyaetal2019,howelletal2019,machadoetal2019,rajaketal2019,machadoetal2020,rubio-abadal2020,pengetal2021,hodsonjarzynski2021}.  As described in Section \ref{diffusion}, the evolution of the driven billiard particle proceeds in three stages: Prethermalization at the initial energy $E_0$, then slow energy absorption and diffusion, and finally the potential breakdown of the rapid driving assumption and the possibility of rapid, unbounded energy absorption. 

However, one characteristic sets the rapidly driven billiard apart from many other Floquet prethermal systems: The $O(\omega^{-2})$ scaling of the energy absorption rate $g_1$. In contrast to this behavior, for a variety of systems subject to rapid periodic driving, previous studies have revealed that prethermal energy absorption rates are \textit{exponentially} small in the driving frequency $\omega$ \cite{abaninetal2015,morietal2016,kuwaharaetal2016,elseetal2017,abaninetal2017,mori2018,howelletal2019,machadoetal2019,tranetal2019,rubio-abadal2020,pengetal2021,hodsonjarzynski2021}. We can understand this discrepancy by reviewing the general account of energy diffusion for Hamiltonian systems under rapid periodic driving, established in \cite{hodsonjarzynski2021}. In this paper, the energy drift and diffusion rates are related to the Fourier transform of an autocorrelation function for the undriven system, evaluated at the drive frequency $\omega$. In \textit{smooth} Hamiltonian systems, this Fourier transform decays faster than any power of $\omega^{-1}$ for large $\omega$ \cite{bracewell1978}, consistent with an energy absorption rate that is exponentially small in $\omega$. However, for a billiard, the discontinuous nature of the collisions with the wall produces a cusp in the relevant autocorrelation function at $t=0$, causing the associated Fourier transform to decay like $\omega^{-2}$. We therefore expect the drift and diffusion coefficients for a billiard to scale like $\omega^{-2}$ for large $\omega$, as verified by the formulas \eqref{g1} and \eqref{g2}.

Our results are also situated in a extensive literature on forced billiards, which have been proposed as models for phenomena ranging from electrical conduction \cite{chernovetal1993,chernovetal2013}, to relativistic charged particle dynamics \cite{veltricarbone2004}, to nuclear dissipation \cite{blockietal1978}. Billiard systems may either be driven via an external force applied between collisions, as in the present paper, or via time-dependence of the billiard walls. In the latter scenario, the billiard boundary is deformed and shifted as a function of time according to a pre-specified schedule, and changes in the particle's energy are induced by collisions with the moving wall. For a variety of models, it has been demonstrated that such systems are susceptible to \textit{Fermi acceleration}: The particle exhibits a  statistical bias towards energy-\textit{increasing} collisions, leading to a systematic growth of the (average) energy \cite{fermi1949,ulam1961,barnettetal2001,karlisetal2006,gelfreichturaev2008,karlisetal2012,batistic2014}. In particular, diffusive energy spreading via this mechanism has been observed for certain models \cite{jarzynski1993,dettmannleonel2013,demersjarzynski2015}.

There is a natural correspondence between billiard systems with time-dependent boundaries and our present model. In Section \ref{diffusion}, we noted that over a single period, the driving force perturbs the velocity of a billiard particle by an amount $\approx \mathbf{F}(\mathbf{x}) \sin (\omega t)/(m \omega)$, and its position by $\approx -\mathbf{F}(\mathbf{x}) \left[\cos (\omega t) - 1 \right]/(m \omega^2)$. Evidently, the particle's motion over a period is well-approximated as small, sinusoidal oscillations about a corresponding undriven trajectory of the particle. Therefore, we can imagine moving to an oscillating reference frame, wherein the particle exhibits approximately undriven motion, and the \textit{walls} perform small, rapid oscillations. Accordingly, we hypothesize that for any rapidly driven billiard satisfying the assumptions of this paper, there is a particular billiard with oscillating boundaries which exhibits energy diffusion with the same drift and diffusion coefficients. For the special case of a standard billiard, $U(\mathbf{x})=0$, we can confirm this correspondence by comparing our results to those of \cite{demersjarzynski2015}, where energy diffusion is established for billiards in the ``quivering limit,'' wherein the walls of a billiard undergo small, rapid periodic oscillations. Under this framework, it is straightforward to verify that if each point on the boundary of a quivering chaotic billiard oscillates about its mean position $\mathbf{x}$ with time-dependence $\mathbf{x} + \mathbf{F}(\mathbf{x}) \cos (\omega t)/m \omega^2$, then the associated drift and diffusion coefficients are exactly those predicted in our model for a standard billiard subject to the force $\mathbf{F}(\mathbf{x})  \cos (\omega t)$. It would be interesting to see whether a similar correspondence is valid in the general case, for $U(\mathbf{x}) \neq 0$.

The results in this paper are also relevant to many-particle systems. To see this, consider a gas of $N$ particles of mass $m$ in a $d$-dimensional billiard cavity, with positions $\mathbf{x}_1 ... \, \mathbf{x}_N$ and velocities $\mathbf{v}_1 ... \, \mathbf{v}_N$. Suppose that these particles interact via some potential $U_{\mathrm{int}} (\{\mathbf{x}_i\})$, and are subject to a driving force $\mathbf{F}_{\mathrm{int}} (\{\mathbf{x}_i\}) \cos (\omega t)$. If we collect the particle positions and velocities into two $(d \times N)$-dimensional vectors $\mathbf{X} \equiv (\mathbf{x}_1 ... \, \mathbf{x}_N)$ and $\mathbf{V} \equiv (\mathbf{v}_1 ... \, \mathbf{v}_N)$, we find that these vectors evolve according to a $(d \times N)$-dimensional version of Newton's law \eqref{newton}. Moreover, when a particle collides with the wall, $\mathbf{V}$ is updated according to a reflection law analogous to \eqref{reflection}, which only reverses the components of $\mathbf{V}$ associated with the colliding particle. Therefore, we see that a many-particle billiard in $d$-dimensional space is mathematically equivalent to a single-particle billiard in $(d \times N)$-dimensional space, where the boundary of the $(d \times N)$-dimensional billiard is given by all points in $\mathbf{X}$-space which correspond to having one or more particles on the $d$-dimensional billiard boundary. This equivalence broadly implies that our results can be extended to many-particle interacting billiards, although the detailed consequences of this equivalence remain to be explored.

Finally, much work has been devoted to understanding energy absorption in periodically driven quantum systems \cite{dalessiorigol2014,lazaridesetal2014,abaninetal2015,ponteetal2015a,rehnetal2016,morietal2016,kuwaharaetal2016,abaninetal2017,notarnicolaetal2018,machadoetal2019,tranetal2019}. It is worth asking how energy diffusion in the classical billiards studied in the present paper might provide insight into the energy dynamics of analogous quantum systems. In accordance with the correspondence principle, we might anticipate that in the semiclassical limit (Planck's constant $h \to 0$), the energy of a rapidly periodically driven, quantized chaotic billiard should evolve diffusively. Indeed, this quantum-classical correspondence may be established \cite{cohen2000,elyutin2006,hodsonjarzynski2021} if one assumes Fermi's golden rule, and invokes semiclassical estimates \cite{feingoldperes1986,wilkinson1987} for the matrix elements of classically chaotic systems. However, it is unclear how to definitively demonstrate such a correspondence starting from unitary quantum dynamics, although much research has been devoted to this problem, particularly with the aid of random matrix theory models \cite{wilkinson1988,wilkinsonaustin1995,cohen2000,cohenkottos2000,elyutin2006}. It may also be fruitful to analyze the quantum analogue of our chaotic billiard with the help of the Floquet-Magnus expansion, which allows the evolution of a system under rapid periodic driving to be expressed perturbatively in powers of $\omega^{-1}$ \cite{bukovetal2015,morietal2018}. It would be interesting to see whether there is a correspondence between our analysis and a Floquet-Magnus approach.

\section*{Acknowledgements}

We acknowledge financial support from the DARPA DRINQS program (D18AC00033), and we thank an anonymous referee for their stimulating and helpful comments.

%TC:ignore

\renewcommand{\theequation}{A\arabic{equation}}
\setcounter{equation}{0}
\setcounter{subsection}{0}

\section*{Appendix A}
\label{appendixA}

Here, we calculate the variance \eqref{varE3} from the expression \eqref{varE0}. We begin by computing the second term in \eqref{varE0}, corresponding to the square of $\langle \Delta \mathcal{ E} \rangle$. Defining the shorthand $a_k \equiv \left( \mathbf{F}_k \cdot \hat{\mathbf{n}}_k \right) \left( \mathbf{v}_k \cdot \hat{\mathbf{n}}_k \right)$, we have:

\begin{equation}
\label{deltaEav}
\langle \Delta \mathcal{E} \rangle = 2 \omega^{-1} \Biggl< \sum_{k=1}^N a_k \sin (\omega t_k) \Biggr> + O(\omega^{-2}).
\end{equation}

The $k^{\mathrm{th}}$ term in this sum depends on $a_k$ and $t_k$, which are ultimately determined by the initial conditions $(\mathbf{x}_0, \mathbf{v}_0)$ for each particle in the ensemble. Since $(\mathbf{x}_0, \mathbf{v}_0)$ is randomly sampled according to the microcanonical distribution \eqref{micro}, $a_k$, $t_k$, and $N$ are all random variables. Therefore, we may express the average of each term as an average with respect to $P_k(a_k,t_k,N)$, the joint probability distribution for $a_k$, $t_k$, and $N$. By the rules of conditional probability, $P_k(a_k,t_k,N)$ may be decomposed as

\begin{equation}
\label{bayes}
P_k(a_k,t_k,N ) = P_k (a_k, N) P_k (t_k | a_k, N),
\end{equation}

\noindent where $P_k (a_k, N )$ is the joint probability distribution for $a_k$ and $N$, and $P_k (t_k | a_k, N)$ is the probability distribution for $t_k$, conditioned on particular values of $a_k$ and $N$. For the $k^{\mathrm{th}}$ term in \eqref{deltaEav}, we then compute the average by summing over all possible values of $N$, and integrating over all values of $a_k$ and $t_k$:

\begin{widetext}
\begin{equation}
\label{deltaEav2}
\langle \Delta \mathcal{E} \rangle = 2 \omega^{-1} \sum_{N=0}^{\infty}  \sum_{k=1}^N \int d a_k \, P_k (a_k,N ) \, a_k \int d t_k \, P_k (t_k | a_k, N) \sin (\omega t_k)  + O(\omega^{-2}).
\end{equation}
\end{widetext}

Recall that the quantities being averaged over are associated with trajectories in an ensemble of \textit{driven} particles. However, for large values of $\omega$, each driven trajectory is only weakly perturbed from its undriven counterpart: The trajectory evolved from the same initial condition $(\mathbf{x}_0, \mathbf{v}_0)$, but with $\mathbf{F} (\mathbf{x}) = \mathbf{0}$. So, to leading order in $\omega^{-1}$, we may replace $P_k (a_k,N)$ with $P_k^0 (a_k,N)$, the joint distribution for $a_k$ and $N$ in the absence of driving. Similarly, we replace $P_k (t_k | a_k, N)$ with $P_k^0 (t_k | a_k, N)$, the conditional distribution for $t_k$ in the absence of driving. Importantly, these new undriven distributions are entirely independent of $\omega$, since they are completely determined by the dynamics of the undriven trajectories. Assuming that these undriven distributions differ from their driven counterparts by an amount of order $O(\omega^{-1})$, we obtain:

\begin{widetext}
\begin{equation}
\label{deltaEav3}
\langle \Delta \mathcal{E} \rangle = 2 \omega^{-1} \sum_{N=0}^{\infty}  \sum_{k=1}^N \int d a_k \, P_k^0 (a_k,N) \, a_k \int d t_k \, P_k^0 (t_k | a_k, N) \sin (\omega t_k)  + O(\omega^{-2}).
\end{equation}
\end{widetext}

Finally, consider the inner integral over $t_k$. The integrand is the product of $P_k^0 (t_k | a_k, N)$, which is independent of $\omega$, and $\sin (\omega t_k)$, an oscillatory function with zero time average. It is straightforward to show that integrals of this form approach zero like $\omega^{-1}$ or faster for large $\omega$. Therefore, this integral is of order $O(\omega^{-1})$ for each $k$, and we are left with

\begin{equation}
\label{deltaEav4}
\langle \Delta \mathcal{E} \rangle =  O(\omega^{-2}).
\end{equation}

\noindent This implies that $\mathrm{Var} (\mathcal{E}) = \langle( \Delta \mathcal{E})^2\rangle +  O(\omega^{-4})$. We may now express \eqref{varE0} as:

\begin{equation}
\label{varE}
\mathrm{Var} (\mathcal{E}) = 4 \omega^{-2} \Biggl<  \sum_{k=1}^N \sum_{l=1}^N  a_k a_l \sin (\omega t_k) \sin (\omega t_l) \Biggr> + O(\omega^{-3}).
\end{equation}

We evaluate this average similarly to how we computed $\langle \Delta \mathcal{E} \rangle$. The logic is the same: To leading order, the average may be calculated with respect to the ensemble of \textit{undriven} trajectories. Then, for $l \neq k$, the integrals over $t_k$ and $t_l$ in the average are of order $O(\omega^{-1})$, because of the oscillating factor $\sin (\omega t_k) \sin(\omega t_l)$ in the integrand. The contribution of the $l \neq k$ terms to $\mathrm{Var} (\mathcal{E})$ is therefore of order $O(\omega^{-3})$, because of the factor $\omega^{-2}$ outside the sum. For the $l = k$ terms, we note that $\sin (\omega t_k) \sin (\omega t_l) = \sin^2 (\omega t_k) = \frac{1}{2} - \frac{1}{2} \cos (2 \omega t_k)$, the sum of a constant term and an oscillatory term. The contributions to $\mathrm{Var} (\mathcal{E})$ corresponding to the oscillatory term $- \frac{1}{2} \cos (2 \omega t_k)$ are also of order $O(\omega^{-3})$. Thus, the only remaining contribution to $\mathrm{Var} (\mathcal{E})$ is given by

\begin{equation}
\label{varE2A}
\mathrm{Var} (\mathcal{E}) = 2 \omega^{-2} \Biggl<  \sum_{k=1}^N  a_k^2 \Biggr>_0 + O(\omega^{-3}),
\end{equation}

\noindent where we have added the subscript $0$ to emphasize that the average is over the ensemble of undriven particles. Recalling that $a_k = \left( \mathbf{F}_k \cdot \hat{\mathbf{n}}_k \right) \left( \mathbf{v}_k \cdot \hat{\mathbf{n}}_k \right)$, we see that this is \eqref{varE2} in the main text.

We briefly pause to interpret this result. In evaluating $\langle( \Delta \mathcal{E})^2\rangle$ and $\langle\Delta \mathcal{E}\rangle$, we have seen that for large $\omega$, the oscillatory factors $\sin (\omega t_k)$ average to zero. These factors become effectively uncorrelated with one another, and with the quantities $a_k$. Intuitively, this lack of correlation arises because otherwise similar trajectories in the ensemble may have totally different values of $\sin (\omega t_k)$: Two nearby trajectories with even a small difference between their associated collision times $t_k$ will have a huge $O(\omega)$ difference in the corresponding values of $\omega t_k$. As a result, the phases $\omega t_k \, \mathrm{mod} \, 2 \pi$ effectively become independent random variables, uniformly distributed on $[0,2 \pi )$.

To reiterate, the average \eqref{varE2A} is taken over a microcanonical ensemble of initial conditions with energy $E_0$, evolved for a time $\Delta t$ according to the \textit{undriven} equations of motion. The sum $\sum_{k=1}^N  a_k^2$ is over all collisions which occur from $t=0$ to $t=\Delta t$. Our strategy will be to decompose this sum into many small contributions, evaluate the average of each contribution, and then add up all these results.

Specifically, let us divide up the boundary of the billiard into infinitesimal patches, indexed by a variable $l$: Each patch is centered on a point $\mathbf{x}^{(l)}$ on the boundary, and has a $(d-1)$-dimensional hyper-area $dS $. Moreover, we partition velocity space into infinitesimal hypercubes of hyper-volume $d^d \mathbf{v}$ labeled by $m$, each centered on a velocity point $\mathbf{v}^{(m)}$. Finally, we divide up the time interval from $t=0$ to $t=\Delta t$ into infinitesimal segments of duration $d t$, beginning at successive times $t^{(n)} = n \,dt = 0,\, dt,\, 2 \,dt \,... $. Let us now sum $a_k^2$, \textit{only} counting collisions associated with a particular choice of the indices $l$, $m$, and $n$: Those collisions which occurred on the patch containing $\mathbf{x}^{(l)}$, with incoming velocity in the velocity cell corresponding to $\mathbf{v}^{(m)}$, between the times $t^{(n)}$ and $t^{(n)}+dt$. If we denote an average over such a restricted sum with $\langle ... \rangle_{0,l,m,n}$, then $\mathrm{Var} (\mathcal{E})$ is just a sum over such averages:

\begin{equation}
\label{X2}
\mathrm{Var} (\mathcal{E}) = 2 \omega^{-2} \sum_{l,m,n} \Biggl<  \sum_{k=1}^N   a_k^2 \Biggr>_{0,l,m,n}  +  O(\omega^{-3}).
\end{equation}

\begin{figure}[ht]
\centering
\includegraphics[width=0.45\textwidth]{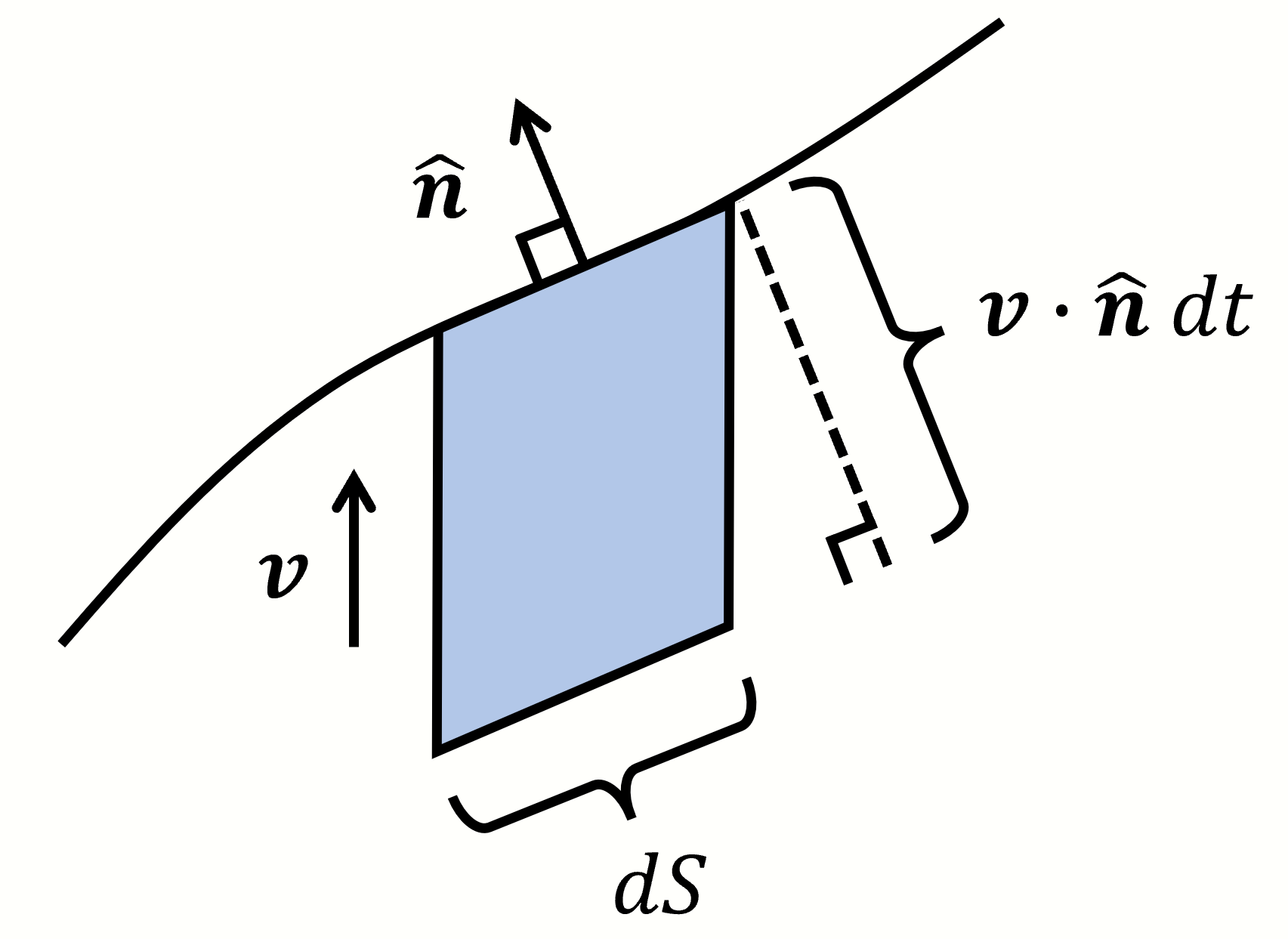}
\caption{Diagram of collisions associated with a given choice of $l$, $m$, and $n$, for the case of $d=2$ dimensions. The curved line represents the billiard boundary, and $\hat{\mathbf{n}} (\mathbf{x}^{(l)}) \equiv \hat{\mathbf{n}}$ is the outward-facing normal near such collisions. Over the infinitesimal time interval from $t^{(n)}$ to $t^{(n)}+ dt$, any particle in the shaded parallelogram with velocity $ \mathbf{v}^{(m)} \equiv \mathbf{v}$ will collide with the boundary sometime during this interval. The area of this parallelogram is $\mathbf{v}\cdot \hat{\mathbf{n}} \,  dt \, dS$, and so collisions associated with $l$, $m$, and $n$ correspond to a phase space volume of $\mathbf{v} \cdot \hat{\mathbf{n}} \, dt \, dS \, d^d \mathbf{v}$. Analogous arguments apply to higher-dimensional billiards.}
\label{fig:parallelogram}
\end{figure}

For a given choice of $l$, $m$, and $n$, what is this average? Well, for all collisions associated with a particular $l$ and $m$, we have that $a_k^2  \approx \left[ \mathbf{F} (\mathbf{x}^{(l)}) \cdot \hat{\mathbf{n}} (\mathbf{x}^{(l)})\right]^2 \left[ \mathbf{v}^{(m)} \cdot \hat{\mathbf{n}} (\mathbf{x}^{(l)})\right]^2 \equiv \left( \mathbf{F}  \cdot \hat{\mathbf{n}} \right)^2 \left( \mathbf{v} \cdot \hat{\mathbf{n}} \right)^2$, so this factor can be brought outside the average. Then, we are simply averaging over the number of collisions corresponding to $l$, $m$, and $n$. This is only nonzero for a small fraction of the ensemble with associated phase space volume $\mathbf{v}^{(m)} \cdot \hat{\mathbf{n}} (\mathbf{x}^{(l)}) \, dt \, dS \, d^d \mathbf{v} \equiv \mathbf{v} \cdot \hat{\mathbf{n}} \,  dt \, dS \, d^d \mathbf{v}$ (see Figure \ref{fig:parallelogram}); the corresponding average is therefore $\rho_{E_0} (\mathbf{x}^{(l)},\mathbf{v}^{(m)}) \equiv \rho_{E_0}$ times this volume elment. Thus, we can convert the sum of over $l$, $m$, and $n$ into an integral over $\mathbf{x}$, $\mathbf{v}$, and $t$, obtaining:

\begin{widetext}
\begin{equation}
\label{X3}
\mathrm{Var} (\mathcal{E}) = 2 \omega^{-2} \Delta t \int dS    \int_{\mathbf{v} \cdot \hat{\mathbf{n}}  > 0 } d^d \mathbf{v} \, \rho_{E_0}   \left( \mathbf{F}  \cdot \hat{\mathbf{n}}  \right)^2 \left( \mathbf{v} \cdot \hat{\mathbf{n}}  \right)^3 +  O(\omega^{-3}).
\end{equation}
\end{widetext}

\noindent Note the restriction to $\mathbf{v} \cdot \hat{\mathbf{n}} (\mathbf{x})> 0$, since a collision can only occur if the incoming velocity $\mathbf{v}$ is directed towards the wall. We can interpret the quantity $\rho_{E_0} (\mathbf{x},\mathbf{v})  \, \mathbf{v} \cdot \hat{\mathbf{n}} (\mathbf{x})  $ as the differential average collision rate in the microcanonical ensemble, for collisions at the point $\mathbf{x}$ on the boundary with incoming velocity $\mathbf{v}$. $\mathrm{Var} (\mathcal{E})$ is then obtained by integrating $\left[ \mathbf{F} (\mathbf{x}) \cdot \hat{\mathbf{n}} (\mathbf{x}) \right]^2 \left[ \mathbf{v} \cdot \hat{\mathbf{n}} (\mathbf{x}) \right]^2$ over all possible collisions, weighted by the rate at which each type of collision occurs.

With the definition of $\rho_{E_0} (\mathbf{x},\mathbf{v})$ (see \eqref{micro}), we may perform the integral over $\mathbf{v}$ using $d$-dimensional spherical coordinates. The result is

\begin{equation}
\label{X4}
 \mathrm{Var} (\mathcal{E}) =  \frac{4 B_{d-1} \omega^{-2} \Delta t}{(d+1) m \Sigma (E_0)} \,   \int dS  \,  v_{E_0}^{d+1} \left( \mathbf{F}  \cdot \hat{\mathbf{n}} \right)^2 + O(\omega^{-3}).
\end{equation}

\noindent Here, $B_{d-1}$ is the hyper-volume of the unit ball in $(d-1)$-dimensional space, and  $v_{E_0} (\mathbf{x}) \equiv v_{E_0}$ is defined as in \eqref{defvE}.

We can rewrite this expression in terms of $\gamma_{E_0} (\mathbf{x})$, the differential average collision rate for collisions at the location $\mathbf{x}$. $\gamma_{E_0} (\mathbf{x})$ is obtained by integrating $\rho_{E_0}  (\mathbf{x},\mathbf{v}) \, \mathbf{v} \cdot \hat{\mathbf{n}} (\mathbf{x}) $ over all $\mathbf{v}$ such that $\mathbf{v} \cdot \hat{\mathbf{n}} (\mathbf{x}) > 0$. This is another spherical integral; the result is given by \eqref{gamma}. Comparing \eqref{gamma} and \eqref{X4}, we obtain \eqref{varE3}.

\renewcommand{\theequation}{B\arabic{equation}}
\setcounter{equation}{0}
\setcounter{subsection}{0}

\section*{Appendix B}
\label{appendixB}

Here, we describe the details of the numerical simulations presented in Section \ref{numerical}. For a particle in the clover billiard subject to a force $\mathbf{F} \cos (\omega t)$, we discuss how to evolve the particle according to the equations of motion, and how to solve the corresponding Fokker-Planck equation.

First, let us describe the evolution of the trajectory ensemble. We consider an ensemble of particles with initial energy $E_0$ at $t=0$, with a microcanonical distribution of initial conditions. For a standard billiard ($U(\mathbf{x})  = 0$), the microcanonical distribution  \eqref{micro} corresponds to sampling the initial positions $\mathbf{x}_0$ from a uniform distribution over the billiard's area, and the initial velocities $\mathbf{v}_0$ from an isotropic distribution with fixed speed $v_0 \equiv \sqrt{2 E_0 /m}$. We generate $N \gg 1$ independent samples in this way, and then evolve each sample in time by alternately integrating the equations of motion \eqref{newton}, and updating the velocity according to the reflection law \eqref{reflection} whenever the particle collides with the wall. In between the $k^{\mathrm{th}}$ and $(k+1)^{\mathrm{th}}$ collisions, we may integrate \eqref{newton} explicitly to obtain $\mathbf{x}_t$ and $\mathbf{v}_t$. Using the same notation as in Section \ref{g1g2}, we find:

\begin{align}
\begin{split}
\label{inbetweenx}
\mathbf{x}_t &= \mathbf{x}_k + \left[ \mathbf{v}_k^+ - \frac{\mathbf{F}}{m \omega} \sin (\omega t_k) \right] (t-t_k) \\ &\quad  - \frac{\mathbf{F}}{m \omega^2} \Big[ \cos (\omega t) - \cos (\omega t_k) \Big],
\end{split}
\end{align}

\begin{equation}
\label{inbetweenv}
\mathbf{v}_t = \mathbf{v}_k^+ + \frac{\mathbf{F}}{m \omega} \Big[ \sin (\omega t) - \sin (\omega t_k) \Big].
\end{equation}

\noindent We see that the particle rapidly oscillates within a small envelope about a straight-line average trajectory. Given the above expressions, finding the next position and velocity at the $(k+1)^{\mathrm{th}}$ collision is simply a matter of solving numerically for where and when this trajectory next intersects with the billiard wall.

In this way, we determine the trajectory of each particle in the ensemble between $t=0$ and some $t=\Delta t$. Then, for any time $t \in [0,\Delta t]$, we compute the energy $\mathcal{E} = \frac{1}{2} m |\mathbf{v}_t|^2$ of each particle, and collect all of these energy values into a histogram. This histogram gives an excellent approximation of the energy distribution $\eta (E,t)$; it only deviates from $\eta (E,t)$ due to the finite number of samples and the small machine error accrued when computing each trajectory.

To compare these results with the energy diffusion description, we then solve the Fokker-Planck equation \eqref{fp}. For a standard billiard, the Fokker-Planck equation admits an analytical solution which has been studied previously. To show this, we note that by \eqref{g1noU} and \eqref{g2noU}, we have $g_1 = C E^{1/2}$ and $g_2 = 4 C E^{3/2}/(d+1)$, where $C$ is a constant independent of energy. If we substitute these expressions into \eqref{fp}, and define the rescaled time variable $s \equiv C t$, then after some manipulations we obtain:

\begin{equation}
\label{fpnoU}
\frac{\partial \eta}{\partial s} = \frac{2}{d+1} \frac{\partial}{\partial E} \left[ E^{(1+d)/2} \frac{\partial}{\partial E} \left( E^{(2-d)/2} \eta \right) \right] .
\end{equation}

\noindent This equation is identical to Equation (60) in \cite{demersjarzynski2015}. This reference also provides the solution to this equation for the initial condition $\eta (E,0) = \delta (E-E_0)$, which we reproduce here:

\begin{widetext}
\begin{equation}
\label{fpsoln}
\eta (E,t) =\eta (E,s/C) = \frac{d+1}{s E_0^{1/2}} \left( \frac{E}{E_0}\right)^{(d-3)/4} I_{d-1} \left[ \frac{4(d+1)}{s} E_0^{1/4} E^{1/4} \right] \exp \left[ - \frac{2 (d+1)}{s} \left( E_0^{1/2} + E^{1/2} \right) \right].
\end{equation}
\end{widetext}

\noindent Here, $I_{d-1} (x)$ is the modified Bessel function of the first kind, of order $d-1$.

It only remains to compute the constant $C$ for the special case of the clover billiard:

\begin{equation}
\label{C}
C = \left( \frac{2}{m} \right)^{3/2} \frac{\omega^{-2}}{\lambda} \frac{1}{S} \int dS \,  \left( \mathbf{F}  \cdot \hat{\mathbf{n}} \right)^2 .
\end{equation}

In $d=2$ dimensions, $S$ is the perimeter of the billiard, and the integral over $dS$ is a line integral over the billiard boundary. For a constant $\mathbf{F} (\mathbf{x}) = \mathbf{F}$, upon performing the appropriate line integrals we find that $S^{-1} \int dS \,  \left[ \mathbf{F} (\mathbf{x}) \cdot \hat{\mathbf{n}} (\mathbf{x})\right]^2 = F^2 /2$, where $F \equiv |\mathbf{F}|$. Then, we can use the relation $\lambda \equiv d \dfrac{B_d}{B_{d-1}} \dfrac{V}{S}$ with $d=2$ to obtain $\lambda = \pi V /S$. In two dimensions, $V$ is the area of the billiard. $V$ and $S$ are geometric quantities which may be computed in terms of the radii $R_1$ and $R_2$. For the specific case of $R_1 = 1$ and $R_2 = 2$, we find that $\lambda \approx 2.610$. Upon combining these results, and setting $m=1$, we obtain:

\begin{equation}
\label{C2}
C \approx 0.5419 \, \omega^{-2} F^2 .
\end{equation}

\noindent With this result, we may now determine the distribution $\eta(E,t)$ at any time $t$, given the parameter choices $m=1$, $R_1 = 1$, $R_2 = 2$, and $\mathbf{F} = F(\hat{\mathbf{x}} + \hat{\mathbf{y}})/\sqrt{2}$. We simply select values for $F$ and $\omega$, and then substitute the resulting value of $C$ into \eqref{fpsoln} (recalling that $s= C t$, and that $d=2$).

%TC:endignore

%apsrev4-2.bst 2019-01-14 (MD) hand-edited version of apsrev4-1.bst
%Control: key (0)
%Control: author (8) initials jnrlst
%Control: editor formatted (1) identically to author
%Control: production of article title (0) allowed
%Control: page (0) single
%Control: year (1) truncated
%Control: production of eprint (0) enabled
%

\end{document}